# Chemical scissors cut phosphorene nanostructures and their novel electronic properties

Xihong Peng, [1*] Qun Wei [1, 2]

[1] School of Letters and Sciences, Arizona State University, Mesa, Arizona 85212, USA

[2] School of Physics and Optoelectronic Engineering, Xidian University, Xi'an, 710071, P.R. China

## ABSTRACT

Phosphorene, a recently fabricated two dimensional puckered honeycomb structure of phosphorus, showed promising properties for applications of nano-electronics. In this work, we report our findings of chemical scissors effects on phosphorene, using first principles density functional theory methods. It was found that several chemical species, such as H, F, Cl and OH group, can act effectively as scissors to cut phosphorene. Phosphorus chains and nanoribbons can be obtained using different surface coverage of the chemical species. The scissor effects of these species are resulted from their strong chemical bonds with the P atoms. Species such as O, S and Se were not able to cut phosphorene nanostructures due to their lack of strong binding with P. The electronic structure calculations of the produced P-chains reveal that the saturated chain is an insulator while the pristine chain demonstrates a Dirac point at X with a Fermi velocity of $8 \times 10^5$ m/s. The obtained zigzag phosphorene nanoribbons show either metallic or semiconducting behaviors, depending on the treatment of the edge P atoms.

**Keywords:**, phosphorene, chemical scissors, phosphorus chains, phosphorene nanoribbons, band structure, density of states, Dirac point, 1D Dirac material, Fermi velocity.

---

[*] Author to whom correspondence should be addressed. Electronic mail: Xihong.peng@asu.edu.

## 1. Introduction

Two dimension (2D) few-layer black phosphorus and phosphorene structures [1-4] has attracted research interests in the society of material science. [5-16] These materials are chemically inert and have shown promising transport properties. They have carrier mobility up to 1000 cm$^2$/V·s and an on/off ratio up to $10^4$ was achieved for the phosphorene transistors at room temperature. [2, 17] Moreover, these materials demonstrate direct band gap at Γ of Brillouin zone [2, 5, 7, 17-19] (in contrast to the zero-gap in graphene), which enable potential applications in optoelectronics.

For practical application, enabling of lithography on 2D layer is essential. Extensive research work has been carried out on graphene. Conventional approaches include e-beam lithography [20] by burning off material, plasma etching, [21] lithography by AFM/STM [22, 23] etc. These methods involve aggressive physical/chemical forces, bringing the concern of introduction of undesired defects into the patterned structures.

Chemical functioning of the surface has been widely explored on graphene for patterning purpose. [24, 25] One example is graphane - hydrogenated graphene. [25-27] In this case, the delocalized π bond of carbon was broken with an introduction of a bond to hydrogen. This hydrogenated surface can be recovered by thermal heating, which makes a promising and controllable way for patterning and tuning the properties of graphene. While hydrogen surface passivation does not break the backbone of C-C bonds and is not a direct method for cutting structures.

However, the situation was found to be different for phosphorene. In this report, the effect of surface passivation on phosphorene was explored using first principles calculations. It was found that some chemical species, for instance H, could dismantle the structure of phosphorene and few-layer black phosphorus into nano patterns, in addition to the traditional role of engineering electronic properties of the materials. The monolayer of black phosphorus has a rippled structure, which can be essentially regarded as the upper and lower half-layer structures. With H passivation the P atoms at the upper half-layer, H and P form a strong σ bond and break the P-P bonds between the upper and lower half-layers of phosphorene. By selecting the location of passivation, the shape and size of phosphorene nanoribbons (PNRs) or chains along the zigzag direction are allowed for a precise control (see the schematic in Figure 1).

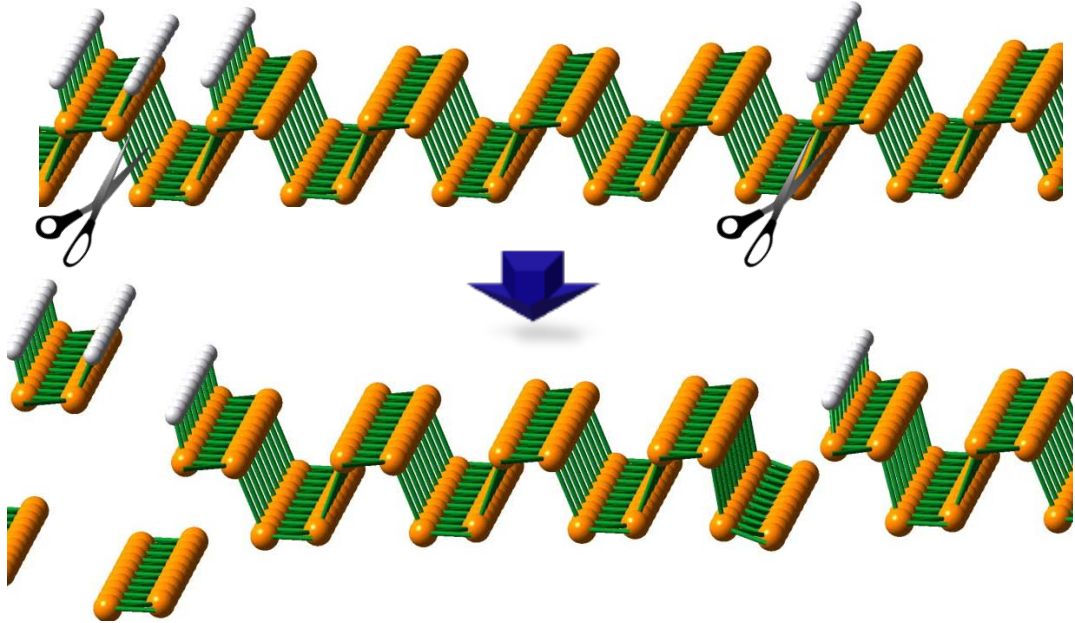

**Figure 1.Schematic of the chemical scissors effects of H on phosphorene.**

For the produced phosphorene zigzag chains, the upper passivated chain has a wide band gap, while the non-passivated counterpart is a one-dimension (1D) Dirac material with the Fermi velocity of $8 \times 10^5$ m/s. This extraordinary electronic propriety of the pristine chain is essentially related to its topographic symmetry. In addition, the obtained pristine nanoribbons show metallic behavior while the both-edge-passivated PNRs are semiconductors. Similarly, other common chemical species, such as F, Cl, hydroxyl group OH, etc., all cause these scissors effects. However, O and S surface passivation do not have this dismantle effect due to their relatively weak unsaturated bonding with P atoms.

## 2. Simulation details

The *ab initio* calculations were carried out using first principles density functional theory (DFT).[28] The Perdew-Burke-Ernzerhof (PBE) exchange-correlation functional [29] and the projector-augmented wave (PAW) potentials [30, 31] were employed. The calculations were performed using the Vienna Ab-initio Simulation Package (VASP).[32, 33] The kinetic energy cutoff for the plane wave basis set was chosen to be 500 eV. The energy convergence criteria for electronic and ionic iterations were set to be $10^{-5}$ eV and $10^{-4}$ eV, respectively. The reciprocal space for the unit cell of monolayer phosphorene was meshed at $14 \times 10 \times 1$ using Monkhorst

Pack meshes centered at Γ point. The non-spin-polarized calculations were used to relax the geometries of the systems. For the phosphorene nanoribbons, spin-polarized calculations were performed to calculate their electronic band structures and density of states (DOS). 21 K-points were included in band structure calculations along Γ to X for the zigzag P-chains and ribbons. A unit cell with periodic boundary condition was used. A vacuum space of at least 20 Å was included in the unit cell to minimize the interaction between the system and its replicas resulting from the periodic boundary condition.

The initial structures of monolayer phosphorene were obtained from bulk black phosphorus.[34] Monolayer black phosphorus has a puckered honeycomb structure with each phosphorus atom covalently bonded with three adjacent atoms. Our calculated lattice constants for bulk black phosphorus are $a = 3.307$ Å, $b = 4.547$ Å, and $c = 11.210$ Å, in good agreement with experimental values [34] and other theoretical calculations.[17, 35] The relaxed lattice constants for a monolayer of phosphorene are $a = 3.295$ Å, $b = 4.618$ Å.

## 3. Results and discussion

### A. Chemical scissors effects

The 2D phosphorene structure is shown in Figure 2. Unlike graphene as a flat structure, phosphorene is puckered. The spatial distance $d_z$ between two half-layer P atoms is 2.10 Å. The bond length $r_1$ and bond angle α within a half-layer (in the xy-plane) are 2.22 Å and 95.9°, respectively. The bond length $r_2$ and bond angle β connecting P atoms on both half-layers are 2.26 Å and 104.1°, respectively.

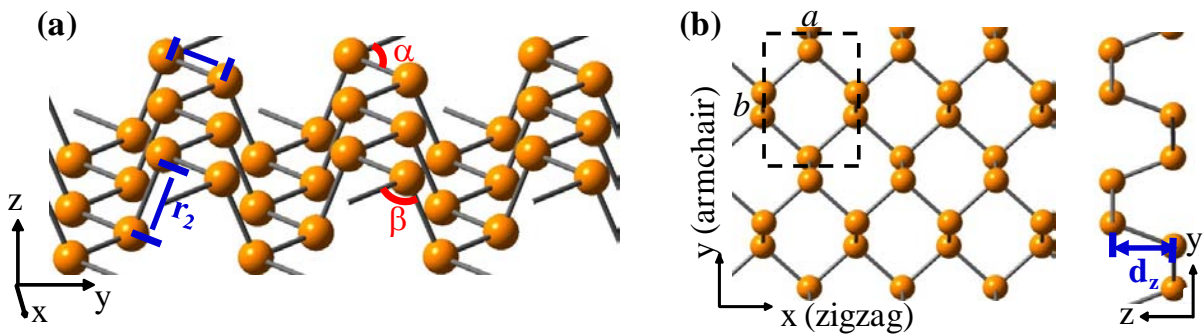

**Figure 2. Snapshots of 2D puckered phosphorene structure. The dashed rectangle in (b) indicates a unit cell.**

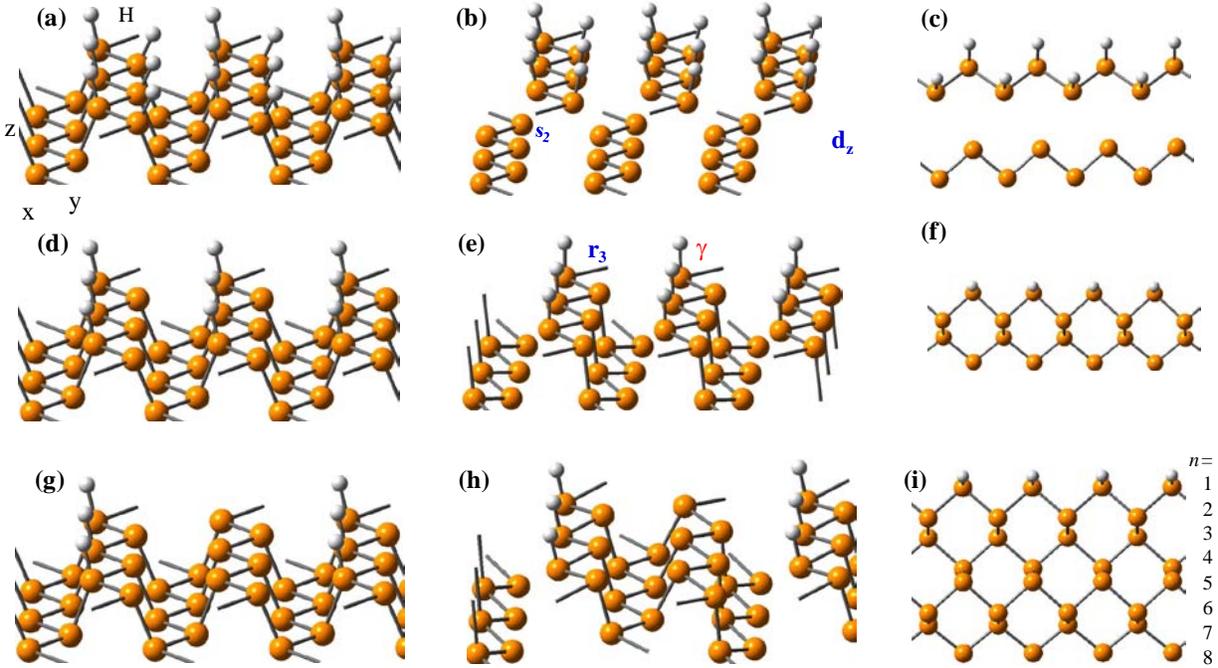

**Figure 3.** Chemical scissors effect of H on phosphorene. (a), (d) and (g) are the initial structures with different surface coverage ratios of H:P (1:1, 1:2, 1:4, respectively). (b), (e) and (h) are the DFT relaxed geometries. The products of the H scissors inside the ovals are displayed in (c) the hydrogenated and pristine zigzag P-chains, (f) the 4L-ZPNR, and (i) the 8L-ZPNR. The H scissors can be used to cut $n$L-ZPNR where $n = 4m$ with $m$ an integer. The orange and white dots represent P and H atoms, respectively.

H atoms were first added on the top half-layer with different surface coverage. For example, each P atom at the top half-layer was terminated with an H atom, which is denoted as surface coverage H:P=1:1, as shown in Figure 3(a). Note that the P atoms at the bottom half-layer were not treated with H, considering the situation that the materials were on a substrate. To simulate a surface coverage smaller than 1, supercells were employed. For instance, a $1 \times 2$ supercell of monolayer phosphorene was used to consider the surface coverage H:P=1:2, in which H atoms were added to 50% of P atoms at the top half-layer (see Figure 3(d)). Similarly, a $1 \times 4$ supercell was used to simulate the surface coverage of 1:4, and so on.

Figure 3 (a),(d), and (g) gives the snapshot of the initial structures with the surface coverage of 1:1, 1:2, and 1:4, respectively. Periodic boundary conditions in the x and y directions and a vacuum space of 20 Å in the z direction were employed and the configuration of the whole system was relaxed within the frame of DFT. Figure 3 (b), (e), and (h) display their relaxed geometries. It is clear that H forms bond with P atoms and effectively breaks the P-P bonds between the top and bottom half-layer of phosphorene. In the case of coverage ratio 1:1 in Figure 3 (b), the full coverage of H on each P atoms breaks all P-P bonds connecting the top and bottom half layers. The products of the H scissors are two types of zigzag P-chains. The top gives a hydrogenated P-chain and the bottom is a pristine P-chain, as shown in Figure 3 (c).

However, for the coverage ratio 1:2 in Figure 3(e), only 50% of P-P bonds between the top and bottom half-layer are broken. This produces a narrow zigzag PNR (ZPNR) with a width of 4.6 Å, in which one edge of the ribbon were saturated with H and the other edge are pristine P atoms. Similarly, for the coverage ratio 1:4 in Figure 3(h), a quarter of P-P bonds were broken by H and results in a relative wider ZPNR with a width of 9.2 Å. The width of a ribbon can also be referred as $n$L, according to the number $n$ of P atoms in the armchair direction (see Figure 3(i)). As an example, Figure 3 (f) and (i) demonstrate the snapshots of 4L-ZPNR and 8L-ZPNR. Furthermore, a coverage ratio of 1:8 will generate a 16L-ZPNR. These H scissors can be used to cut $n$L-ZPNR (where $n = 4m$ with $m$ an integer) with a width $w = m * 4.6$ Å.

The bond lengths and angles for the geometrically relaxed structures with different H surface coverage ratios are listed in Table I. Compared to that of monolayer phosphorene, The bond length $r_1$ and bond angle $\alpha$ within a half-layer (in the xy-plane) have negligible changes. However, the bond length $r_2$ (or P-P distance $s_2$) and the spatial distance $d_z$ between two half-layer P atoms show significant increases due to the scissor effect.

In addition to H, we investigated other chemical species, such as F, Cl, OH group, O, S and Se. It was found that F, Cl and OH groups demonstrate a similar scissors effect as H. However, adding O, S, and Se on the surface of phosphorene did not break the P-P bonds connecting the top and bottom half-layers. As an example, Figure 4 displays their relaxed geometries with the surface coverage of 1:2. For the O (S or Se) case, we checked different initial configurations. For instance, one is to put O (S or Se) near the top of one P atom. Another is to place O (S or Se) on the bridge of two P atoms. The former is energetically more favorable compared to the latter. The relaxed geometries of all configurations show that O (S or Se) did not break the P-P bonds.

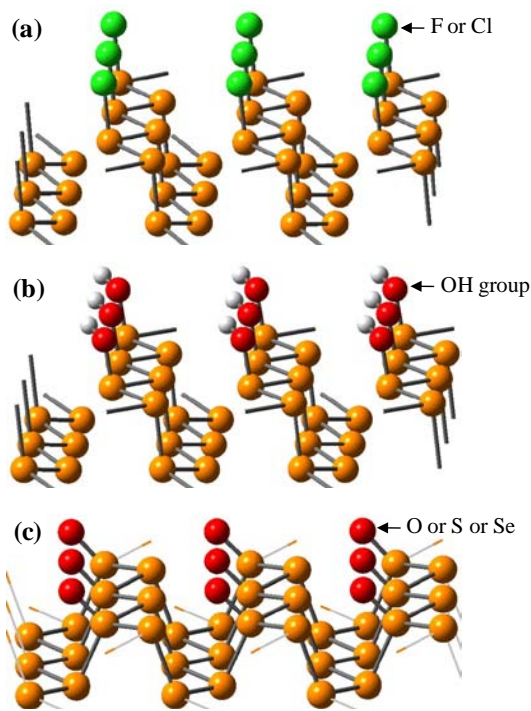

**Figure 4.** The snapshots of the DFT relaxed geometries of additional (a) (b) successful and (c) failed chemical scissors for phosphorene.

**Table 1.** The relaxed bond lengths and angles for the studied chemical species on monolayer phosphorene. The bond lengths and angles were denoted in Figures 1 and 2. As a reference, the bond lengths/angles in monolayer phosphorene were also listed.

| system | surface coverage | $r_1$ (Å) | $r_2$ or $s_2$ (Å) | $d_z$ (Å) | α (°) | β (°) | γ (°) |
|---|---|---|---|---|---|---|---|
| monolayer | 0 | 2.22 | 2.26 | 2.10 | 95.9 | 104.1 | n/a |
| H:P | 1:1 | 2.23 | 3.55 | 3.43 | 95.5 | n/a | 98.6 |
|  | 1:2 | 2.25 | 3.29 | 2.87 | 94.4 | 96.5 | 93.2 |
|  | 1:4 | 2.26 | 3.82 | 3.35 | 93.9 | 93.9 | 92.0 |
|  | 1:6 | 2.26 | 3.83 | 3.36 | 93.9 | 93.0 | 92.2 |
|  | 1:8 | 2.26 | 3.86 | 3.38 | 93.8 | 92.6 | 91.9 |
|  | 1:10 | 2.25 | 3.61 | 3.19 | 94.0 | 93.3 | 91.7 |
| F:P | 1:6 | 2.27 | 3.38 | 3.15 | 93.2 | 96.0 | 96.0 |
| Cl:P | 1:6 | 2.27 | 3.47 | 3.19 | 93.3 | 95.5 | 96.1 |
| OH:P | 1:6 | 2.26 | 3.43 | 3.05 | 93.1 | 92.8 | 94.7 |
| O:P | 1:6 | 2.24 | 2.24 | 2.03 | 95.0 | 98.9 | 112.6 |
| S:P | 1:6 | 2.24 | 2.25 | 2.03 | 94.7 | 100.1 | 112.7 |
| Se:P | 1:6 | 2.25 | 2.27 | 1.94 | 94.4 | 98.6 | 109.9 |

Table I also list the geometry parameters such as bond lengths and angles for these additional species. Similar to H, the bond distance $r_2$ (or $s_2$) and the spatial distance $d_z$ are largely increased for F, Cl and OH, when compared to that of monolayer phosphorene. However, in the cases of O, S and Se, $r_2$ and $d_z$ are very close to that of monolayer. It is worth to mention that the bond angle γ (denoted in Figure 3) has similar value (92° ~ 99°) for the chemical species which show scissors effect. However, for O, S and Se, the bond angle γ increases by about 20°. This is resulted from their different structural geometries and electronic orbital orientations. As shown in Figure 3(a), the P atoms form a bond with F (Cl) and its bond with P at the bottom half-layer is broken. This makes the P atoms still have three covalent bonds, among which the bond angles α, β, and γ are nearly 90° resulted from their particular p-orbital orientations ($p_x$, $p_y$, and $p_z$). On the other hand in Figure 4 (c), those P atoms form four bonds; three are the original bonds with P and one additional with O (S or Se). This makes P in the nearly tetrahedral configurations and increased the bond angle γ to 112°.

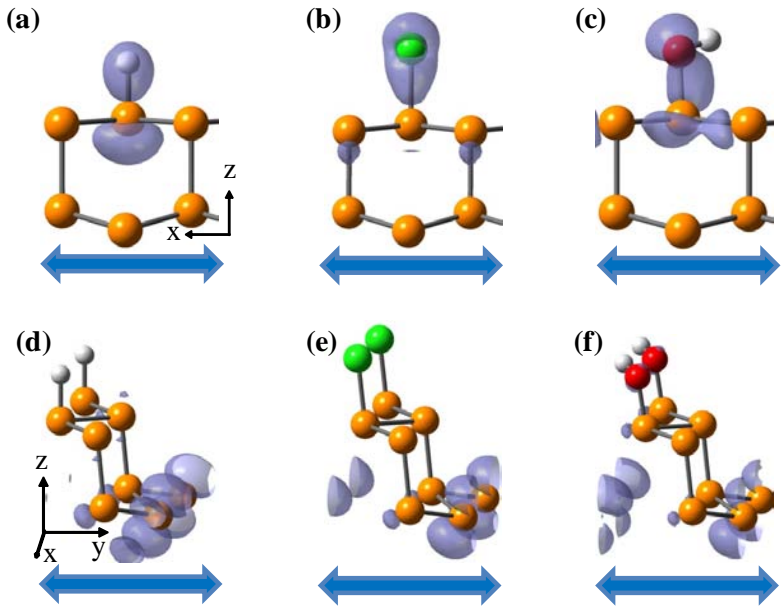

**Figure 5.** The electron density contour plots of the (a) P-H (b) P-Cl (c) P-O bond for the chemical species of H, Cl, and OH group, respectively. The scissor effects are resulted from their strong bonding with the P atoms. (d)-(f) The electron density of the near-Fermi-level states contributed by the edge P atoms at the pristine side in each case. The horizontal blue arrows indicate the periodic boundary directions of the simulation cells. The orange, white, blue and red dots represent P, H, Cl, and O atoms, respectively.

To further understand the chemical scissors effect, especially, why some chemical species work while other not, we examined in detail the wavefunction and charge density of the electronic orbitals of the systems. It was found that whether a chemical species can effectively cut phosphorene depends on if it is strongly bonded to the P atoms. For the cases of H, F, Cl and OH group, these species form a strong bond with P. For example, s-orbital of H is bonded with the $p_z$-orbital of P. Figure 5(a) plots its electron density contour plot of this structure as shown in Figure 3(e) with both periodic boundaries in the x and y directions. The orbital energy of this H-P bond is 5.33 eV below the Fermi level. Figure 5(b) shows a strong $p_z$-orbital binding between Cl and P for the structure shown in Figure 4(a) and its orbital energy is 5.23 eV below the Fermi energy. Figure 5(c) displays the $p_z$-bond between P and O in the OH group for the structure displayed in figure 4(b), and the orbital energy is 6.08 eV under the Fermi level. However, for the cases of O, S, and Se shown in Figure 4(c), we could not found such strong bond between P and these species.

## B. Electronic properties of P-chains and ribbons

We further calculated the electronic properties of the products (i.e. P-chains and nanoribbons) generated from the chemical scissors. Figure 6 presents the band structure and DOS of the two P-chains. Left is the hydrogenated and right is the pristine zigzag P-chains. The hydrogenated chain is an insulator and the DFT predicted band gap is 3.82 eV. For other chemical species such as Cl, and OH, the DFT gaps are 2.78 eV and 2.28 eV, respectively.

However, the pristine P-chain in Figure 6(b) shows zero gap at X. More interestingly, the dispersion curve of the conduction and valence bands near the X point demonstrate a linear relation, which makes the X k-vector a Dirac point for this structure. We used the hybrid functional HSE method [36, 37] to check the band structure of the pristine P-chain and found that the Dirac point still preserves at X. This linear dispersion at X was also observed using the advanced GW calculations [38, 39] although the GW predicted a slightly opened gap of 0.02 eV. The linear dispersion relation at X can be expressed as $E = \pm \beta k$, where $k$ is the distance of wave vector away from the point X and the slope $\beta$ can be fitted from the first principles calculation. According to the relativistic energy-momentum relationship $E = \sqrt{m_0^2 c^4 + p^2 c^2}$, the linear dispersion relation at X implies that the charge carriers are massless fermions traveling at an effective speed of light $v$ (Fermi velocity): $E = vp = v\hbar K = v\hbar k 2\pi / a = vhk/a$, where $a$ is the lattice constant of the pristine zigzag P-chain. Therefore, the Fermi velocity $v$ can be calculated from the expression $v = \beta a / h$. Our calculated Fermi velocity for the pristine P-chain is ~ $8 \times 10^5$ m/s.

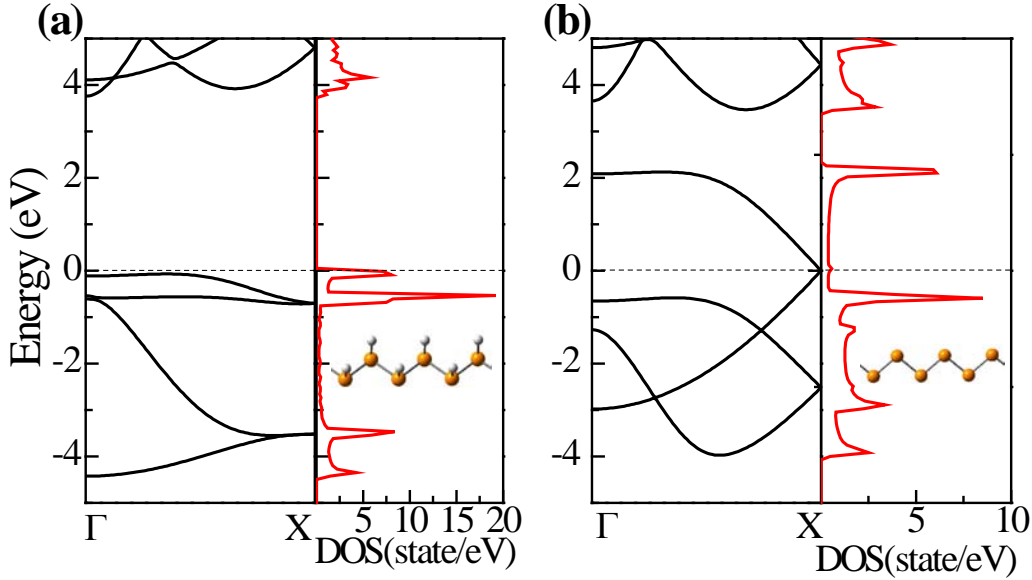

**Figure 6.** Band structure and density of states of the (a) hydrogenated and (b) pristine zigzag P-chains. The hydrogenated P-chain is an insulator with the DFT predicted gap of 3.82 eV. The pristine P-chain has a zero band gap and demonstrates a Dirac point at X. The Fermi level is set at the valence band maximum which is aligned at zero. The insets are the snapshots of the corresponding P-chains.

Figure 7 displays the DOS for the ZPNRs with different widths, namely 4L-ZPNR ($w = 4.6$ Å) and 12L-ZPNR ($w = 13.8$Å). Figure 7 (a) and (b) are their corresponding plots of the DOS. Both show metallic behavior. The bands cross the Fermi level are contributed by the edge P atoms on the pristine side which is clearly demonstrated by the electron density plot in Figure 5(d). Adding H to those P atoms on the pristine side will open the band gap. [40] Figure 7(c) and (d) are the DOS for the 4L-PNR and 12L-ZPNR with both edges saturated using H. The ribbons are semiconductors and the band gap reduces with the increase of the ribbon width due to quantum confinement.

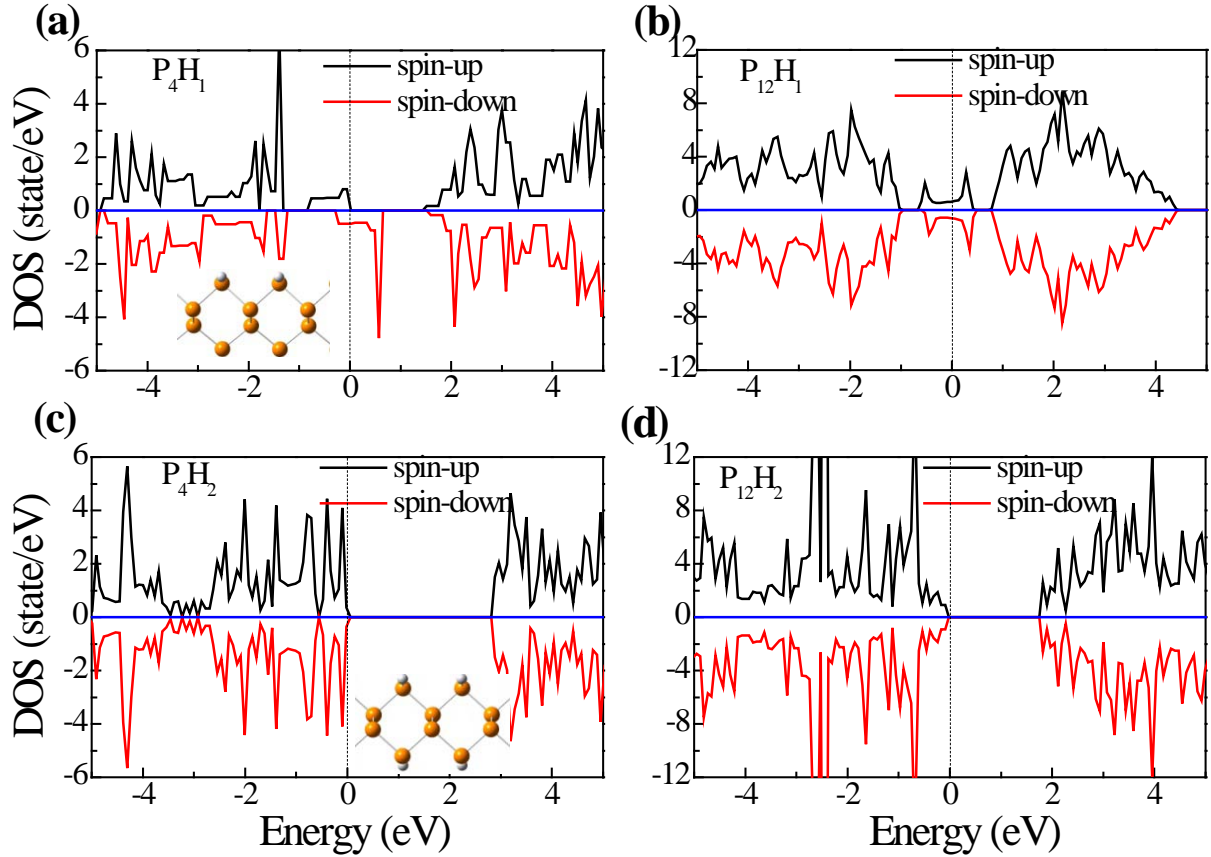

**Figure 7.** DOS of the 4L-ZPNR with (a) one edge passivated with H and the other edge pristine, (b) both edges saturated with H. DOS of the 12L-ZPNR with (c) one edge passivated with H and the other edge pristine, (d) both edges saturated with H. The Fermi level is set at zero. The insets in (a) and (c) are the snapshots of the corresponding 4L-ZPNRs.

## 4. Conclusion

We employed first principles DFT calculations to study the chemical scissors effects on phosphorene. We found that several chemical species such as H, F, Cl and OH group can be used as chemical scissors to cut phosphorene, while others O, S and Se cannot, depending on if the species forms a strong chemical bond with P. Phosphorus chains and nanoribbons can be produced using these chemical scissors. The saturated P-chain is insulating, while the pristine P-chain is a one-dimensional Dirac material with the Fermi velocity of $8 \times 10^5$ m/s. The zigzag nanoribbons are either metallic or semiconductor, depending on the treatment of the edge P atoms. The ribbons are semiconductors if the dangling bonds of the edge P atoms are saturated. They show metallic behavior if dangling bonds exists.

**Acknowledgement**

This work is supported by the Faculty Research Fund from School of Letters and Sciences at Arizona State University (ASU) to Peng. The authors thank ASU Advanced Computing Center and XSEDE for providing computing resources. Dr. F. Tang is acknowledged for the helpful discussions and we thank A. Copple for his critical review of the manuscript.

1. **H. Liu, A. T. Neal, Z. Zhu, Z. Luo, X. Xu, D. Tománek, and P. D. Ye, (2014), Phosphorene: An Unexplored 2D Semiconductor with a High Hole Mobility,** *Acs Nano* **8, 4033-4041.**
2. **Likai Li, Yijun Yu, Guo Jun Ye, Qingqin Ge, Xuedong Ou, Hua Wu, Donglai Feng, Xian Hui Chen, and Yuanbo Zhang, (2014), Black phosphorus field-effect transistors,** *arXiv:1401.4117*
3. **F. Xia, H. Wang, and Y. and Jia, (2014), Rediscovering Black Phosphorus: A Unique Anisotropic 2D Material for Optoelectronics and Electronics,** *arXiv:1402.0270*
4. **E. S. Reich, (2014), Phosphorene excites materials scientists,** *Nature* **506, 19.**
5. **Vy Tran, Ryan Soklaski, Yufeng Liang, and Li Yang, (2014), Layer-Controlled Band Gap and Anisotropic Excitons in Phosphorene and Few-Layer Black Phosphorus,** *arXiv:1402.4192*
6. **Ruixiang Fei and Li Yang, (2014), Strain-Engineering Anisotropic Electrical Conductance of Phosphorene and Few-Layer Black Phosphorus,** *Nano Lett.***, DOI: 10.1021/nl500935z.**
7. **X. Peng, A. Copple, and Q. Wei, (2014), Strain engineered direct-indirect band gap transition and its mechanism in 2D phosphorene,** *arXiv:1403.3771*
8. **J. Dai and X. C. Zeng, (2014), Bilayer Phosphorene: Effect of Stacking Order on Bandgap and its Potential Applications in Thin-Film Solar Cells,** *The Journal of Physical Chemistry Letters* **5, 1289-1293.**
9. **H. Guo, N. Lu, J. Dai, X. Wu, and X. C. Zeng, (2014), Phosphorene nanoribbons, nanotubes and van der Waals multilayers,** *arXiv:1403.6209*
10. **Q. Wei and X. Peng, (2014), Superior mechanical flexibility of phosphorene and few-layer black phosphorus,** *arXiv:1403.7882*
11. **W. Lu, H. Nan, J. Hong, Y. Chen, C. Zhu, Z. Liang, X. Ma, Z. Ni, C. Jin, and Z. Zhang, (2014), Plasma-assisted fabrication of monolayer phosphorene and its Raman characterization,** *Nano Research***, DOI:10.1007/s12274-12014-10446-12277.**
12. **V. Tran and L. Yang, (2014), Unusual Scaling Laws of the Band Gap and Optical Absorption of Phosphorene Nanoribbons,** *arXiv:1404.2247*
13. **A. Maity, A. Singh, and P. Sen, (2014), Peierls transition and edge reconstruction in phosphorene nanoribbons,** *arXiv:1404.2469*
14. **A. Carvalho, A. S. Rodin, and A. H. C. Neto, (2014), Phosphorene nanoribbons,** *arXiv:1404.5115*
15. **H. Y. Lv, W. J. Lu, D. F. Shao, and Y. P. Sun, (2014), Large thermoelectric power factors in black phosphorus and phosphorene,** *arXiv:1404.5171*
16. **Michele Buscema, Dirk J. Groenendijk, Sofya I. Blanter, Gary A. Steele, Herre S.J. van der Zant, and A. Castellanos-Gomez, (2014), Fast and broadband photoresponse of few-layer black phosphorus field-effect transistors,** *arXiv:1403.0565*
17. **Han Liu, Adam T. Neal, Zhen Zhu, David Tomanek, and Peide D. Ye, (2014), Phosphorene: A New 2D Material with High Carrier Mobility,** *arXiv:1401.4133*
18. **Y. Takao and A. Morita, (1981), Electronic structure of black phosphorus: Tight binding approach,** *Physica B & C* **105, 93-98.**
19. **A. S. Rodin, A. Carvalho, and A. H. Castro Neto, (2014), Strain-induced gap modification in black phosphorus,** *arXiv:1401.1801*

20. M. D. Fischbein and M. Drndić, (2008), Electron beam nanosculpting of suspended graphene sheets, *Applied Physics Letters* **93**, 113107.
21. J. Bai, X. Duan, and Y. Huang, (2009), Rational Fabrication of Graphene Nanoribbons Using a Nanowire Etch Mask, *Nano Letters* **9**, 2083-2087.
22. L. Tapaszto, G. Dobrik, P. Lambin, and L. P. Biro, (2008), Tailoring the atomic structure of graphene nanoribbons by scanning tunnelling microscope lithography, *Nature Nanotechnology* **3**, 397-401.
23. L. Weng, L. Zhang, Y. P. Chen, and L. P. Rokhinson, (2008), Atomic force microscope local oxidation nanolithography of graphene, *Applied Physics Letters* **93**, 093107.
24. L. Zhang, S. Diao, Y. Nie, K. Yan, N. Liu, B. Dai, Q. Xie, A. Reina, J. Kong, and Z. Liu, (2011), Photocatalytic Patterning and Modification of Graphene, *Journal of the American Chemical Society* **133**, 2706-2713.
25. J. Feng, W. Li, X. Qian, J. Qi, L. Qi, and J. Li, (2012), Patterning of graphene, *Nanoscale* **4**, 4883-4899.
26. D. C. Elias, R. R. Nair, T. M. G. Mohiuddin, S. V. Morozov, P. Blake, M. P. Halsall, A. C. Ferrari, D. W. Boukhvalov, M. I. Katsnelson, A. K. Geim, and K. S. Novoselov, (2009), Control of Graphene's Properties by Reversible Hydrogenation: Evidence for Graphane, *Science* **323**, 610-613.
27. P. Sessi, J. R. Guest, M. Bode, and N. P. Guisinger, (2009), Patterning Graphene at the Nanometer Scale via Hydrogen Desorption, *Nano Letters* **9**, 4343-4347.
28. W. Kohn and L. J. Sham, (1965), Self-consistent equations including exchange and correlation effects, *Physical Review* **140**, A1133-A1138.
29. J. P. Perdew, K. Burke, and M. Ernzerhof, (1996), Generalized Gradient Approximation Made Simple, *Physical Review Letters* **77**, 3865-3868.
30. P. E. Blochl, (1994), Projector augmented-wave method, *Physical Review B* **50**, 17953-17979.
31. G. Kresse and D. Joubert, (1999), From ultrasoft pseudopotentials to the projector augmented-wave method, *Physical Review B* **59**, 1758-1775.
32. G. Kresse and J. Furthmuller, (1996), Efficient iterative schemes for ab initio total-energy calculations using a plane-wave basis set, *Physical Review B* **54**, 11169.
33. G. Kresse and J. Furthmuller, (1996), Efficiency of ab-initio total energy calculations for metals and semiconductors using a plane-wave basis set, *Computational Materials Science* **6**, 15-50.
34. Allan Brown and Stig Rundqvist, (1965), Refinement of the crystal structure of black phosphorus, *Acta Cryst.* **19**, 684.
35. Jingsi Qiao, Xianghua Kong, Zhi-Xin Hu, Feng Yang, and Wei Ji, (2014), Few-layer black phosphorus: emerging 2D semiconductor with high anisotropic carrier mobility and linear dichroism, *arXiv:1401.5045*
36. J. Heyd, G. E. Scuseria, and M. Ernzerhof, (2003), Hybrid functionals based on a screened Coulomb potential, *The Journal of Chemical Physics* **118**, 8207-8215.
37. J. Heyd, G. E. Scuseria, and M. Ernzerhof, (2006), Erratum: ``Hybrid functionals based on a screened Coulomb potential'' [J. Chem. Phys. 118, 8207 (2003)], *The Journal of Chemical Physics* **124**, 219906-219901.
38. L. Hedin, (1965), New method for calculating the one-particle Green's function with application to the electron-gas problem, *Physical Review* **139**, A796-A823.

39. M. Shishkin and G. Kresse, (2006), Implementation and performance of the frequency-dependent GW method within the PAW framework, *Physical Review B* **74, 035101.**
40. X. Peng, Q. Wei, and A. Copple, (2014), Edge effects on the electronic properties of phosphorene nanoribbons, *arXiv:1404.5995*